\documentclass[9pt,twocolumn]{article}
\usepackage{times}
\usepackage{url}
\usepackage{graphicx}
\usepackage{amsmath}

\begin{document}
\title{Trends in Social Media : Persistence and Decay}

\author{Sitaram Asur\thanks{Social Computing Lab, HP Labs, Palo Alto, California, USA} \and Bernardo A. Huberman\footnotemark[1]
\and Gabor Szabo\footnotemark[1] \and Chunyan Wang\thanks{Dept of Applied Physics, Stanford University, California, USA}}
\date{\today}

\maketitle

\begin{abstract}

Social media generates a prodigious wealth of real-time content at an incessant rate. From all the content that 
people create and share, only a few topics manage to attract enough attention to rise to the top and become temporal trends which are displayed to users. 
The question of what factors cause the formation and persistence of trends is an important one that
has not been answered yet. In this paper, we conduct an intensive study of trending topics on Twitter and provide a theoretical 
basis for the formation, persistence and decay of trends. We also demonstrate empirically
how factors such as user activity and number of followers do not contribute strongly to trend creation and its propagation. In fact, we find that the resonance of the content with the users of the social network plays a major role in causing trends. 

\end{abstract}

\section{Introduction}

Social media is growing at an explosive rate, with millions of people all over the world 
generating and sharing  content on a scale barely imaginable a few years
ago. This has resulted in massive participation with countless number of updates, opinions, news, comments and 
product reviews being constantly posted and discussed in social web sites such as Facebook, Digg and Twitter, to name
a few. 

This widespread generation and consumption of content has created an extremely competitive online environment where 
different types of content vie with each other for the scarce attention of the user community.
In spite of the seemingly chaotic fashion with which all these interactions take place, certain
topics manage to attract an inordinate amount of attention, thus bubbling to the top
in terms of popularity. Through their visibility, this popular topics contribute to the collective awareness of what is trending and at times can also  affect the public agenda of
the community. 

At present there is no clear picture of  what causes these topics to become extremely popular, nor how some persist in the public eye longer 
than others. There is considerable evidence that one aspect that causes topics to decay over time is their novelty~\cite{Wu07}. Another factor responsible for their decay is the competitive nature of the medium. As  content starts propagating throught a social network it can usurp 
the positions of earlier topics of interest, and due to the limited attention of users it is soon 
rendered invisible by newer content. Yet another aspect responsible for the popularity of certain topics  is the influence of members of the network on the propagation 
of content. Some users generate content that resonates very strongly with their followers thus causing the content to propagate and
gain popularity~\cite{Romero2011}.

The source of that content can originate in standard media outlets or from users who generate topics that eventually become part of the trends and capture the attention of large communities. In either case the fact that a small set of topics become part of the trending set means that they will capture the attention of a large audience for a  short time, thus contributing in some measure to the public agenda. When topics originate in media outlets, the social medium acts as filter and amplifier of what the standard media produces and thus contributes to the agenda setting mechanisms that have been thoroughly studied for more than three decades~\cite{McCombs1993} .

In this paper, we study trending topics on Twitter, an immensely popular microblogging network on which 
millions of users create and propagate enormous content via a steady stream on a daily basis. The trending topics, which 
are shown on the main website, represent those pieces of content that bubble to the surface on Twitter owing to 
frequent mentions by the community. Thus they can be equated to crowdsourced popularity.
We then determine the factors that contribute to the creation and evolution of these trends, as they provide 
insight into the complex interactions that lead to the popularity and persistence of certain topics on Twitter, while most others 
fail to catch on and are lost in the flow.

We first analyze the distribution of the number of tweets across trending topics. We observe that they are characterized by a strong log-normal distribution, 
similar to that found in other networks such as Digg and which is generated by a stochastic multiplicative process~\cite{Wu07}. We also find that the decay function for the tweets is mostly linear. Subsequently we study the persistence of the trends to determine which topics last long at the top. Our analysis reveals that there are few topics that last for long times, while most topics break fairly quickly, in the order of 20-40 minutes. 
Finally, we look at the impact of users on trend persistence times within Twitter.  We find that traditional notions of user influence such as the frequency of posting and the number of followers are not the main drivers of trends, as previously thought. Rather, long trends are characterized by the resonating nature of the content, which is found to arise mainly from traditional media sources. We observe that social media behaves as a selective amplifier for the content generated by traditional media, with chains of retweets by many users leading to the observed trends.

\section{Related work}\label{related}

There has been some prior work on analyzing connections on Twitter. Huberman et al.~\cite{Huberman2008Social} studied social interactions on Twitter to
reveal that the driving process for usage is a sparse hidden network underlying the
friends and followers, while most of the links represent meaningless
interactions. Jansen et al.~\cite{Jansen2009Twitter} have examined Twitter as a
mechanism for word-of-mouth advertising. They considered particular brands and
products and examined the structure of the postings and the change in
sentiments. Galuba et al.~\cite{galuba10retweets}  proposed a propagation model
that predicts which users will tweet about which URL based on the history of past
user activity.

Yang and Leskovec~\cite{Yang2011}  examined patterns of temporal
behavior for hashtags in Twitter. They presented a stable time series clustering
algorithm and demonstrate the common temporal patterns that tweets containing
hashtags follow. There have also been earlier studies focused on social
influence and propagation. Agarwal et al.~\cite{Agarwal2008Identifying} 
studied the problem of identifying influential bloggers in the blogosphere.
They discovered that the most influential bloggers were not necessarily the most
active. Aral et al,~\cite{aral09} have distinguished the effects of homophily
from influence as motivators for propagation. As to the study of influence
within Twitter, Cha et al.~\cite{Cha2010Measuring}  performed a comparison
of three different measures of influence - indegree, retweets, and user mentions.
They discovered that while retweets and mentions correlated well with each
other, the indegree of users did not correlate well with the other two measures.
Based on this, they hypothesized that the number of followers may not a good
measure of influence. Recently, Romero and others~\cite{Romero2011}  introduced 
a novel influence measure that takes into account the passivity of the audience in the social network. They developed 
an iterative algorithm to compute influence in the style of the HITS algorithm and empirically demonstrated  that the number of followers is a poor measure of influence.

\section{Twitter}
Twitter is an extremely popular online microblogging service, that has gained a
very large user following, consisting of close to 200 million users. The Twitter graph 
is a directed social network, where each user chooses
to follow certain other users. Each user submits periodic status updates, known
as \emph{tweets}, that consist of short messages limited in size to 140
characters. These updates typically consist of personal information about the
users, news or links to content such as images, video and articles. The posts
made by a user are automatically displayed on the user's profile page, as well
as shown to his followers. A \emph{retweet} is a post originally made by one user that is forwarded by
another user. Retweets are useful for propagating interesting posts and links
through the Twitter community.

Twitter has attracted lots of attention from corporations due to the immense
potential it provides for viral marketing. Due to its huge reach, Twitter is
increasingly used by news organizations to disseminate news updates, which are
then filtered and commented on by the Twitter community. A number of businesses
and organizations are using Twitter or similar micro-blogging services to
advertise products and disseminate information to stockholders.

\section{Twitter Trends Data}

\emph{Trending topics} are presented as a list by Twitter on their main Twitter.com site, and are selected by an
algorithm proprietary to the service. They mostly consist of two to three word
expressions, and we can assume with a high confidence that they are snippets
that appear more frequently in the most recent stream of tweets than one would
expect from a document term frequency analysis such as TFIDF. The list of trending topics is updated every few minutes as new topics 
become popular. 

Twitter provides a Search API for extracting tweets containing particular
keywords. To obtain the dataset of trends for this study, we repeatedly used the
API in two stages. First, we collected the trending topics by doing an API query
every 20 minutes. 
Second, for each trending topic, we used the Search API to
collect all the tweets mentioning this topic over the past 20 minutes. For each
tweet, we collected the author, the text of the tweet and the time it was
posted. Using this procedure for data collection, we obtained 16.32 million
tweets on 3361 different topics over a course of 40 days in Sep-Oct 2010.

We picked 20 minutes as the duration of a timestamp after evaluating different time lengths, to optimize the 
discovery of new trends while still capturing all trends. This is due to the fact that Twitter only allows 1500 tweets per search query. We found that with 20 minute intervals, we were able to capture all the tweets for the trending topics efficiently.

We noticed that many topics become trends again after they stop trending
according to the Twitter trend algorithm.  We therefore considered these trends as separate sequences: it is very likely
that the spreading mechanism of trends has a strong time component with an
initial increase and a trailing decline, and once a topic stops trending, it
should be considered as new when it reappears among the users that become aware
of it later. This procedure split the 3468 originally collected trend titles
into 6084 individual trend sequences.

\section{Distribution of tweets}
\begin{figure*}[hbtp]
(a) \includegraphics[width=\linewidth]{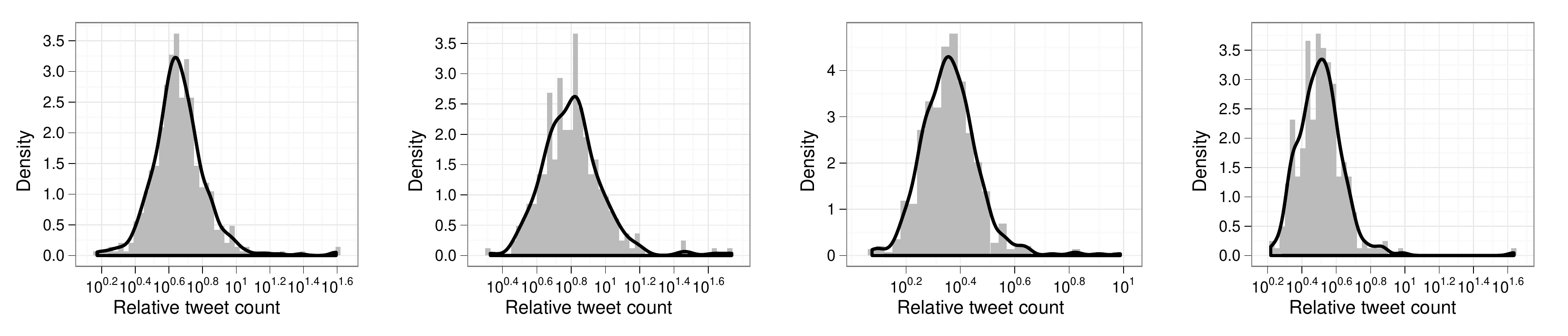}
(b) \includegraphics[width=\linewidth]{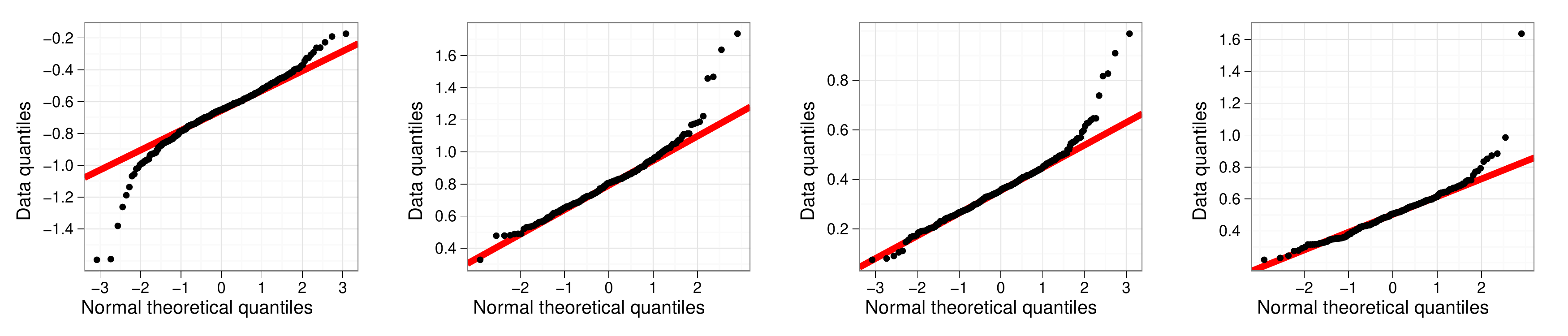}
\vspace{-0.1in}
\caption{\bf{(a) The densities of the ratios between cumulative tweet counts
measured in two respective time frames. From left to right in the figure, the
indices of the time frames between which the ratios were taken are: (2, 10), (2,
14), (4, 10), and (4, 14), respectively. The horizontal axis has been rescaled
logarithmically, and the solid line in the plots shows the density estimates
using a kernel smoother. \emph{(b)} The Q-Q plots of the cumulative tweet
distributions with respect to normal
distributions. If the random variables of the data were a linear transformation
of normal variates, the points would line up on the straight lines shown in the
plots. The tails of the empirical distributions are apparently heavier than in
the normal case.}}
\label{fig:count_densities}
\end{figure*}
We measured the number of tweets that each topic gets in 20 minute intervals,
from the time the topic starts trending until it stops, as described earlier. From this we can sum up the 
tweet counts over time to obtain the cumulative number of tweets $N_q(t_i)$ of topic $q$ for any time
frame $t_i$,
\begin{equation}
N_q(t_i) = \sum_{\tau = 1}^{i} n_q(t_\tau),
\end{equation}
where $n_q(t)$ is the number of tweets on topic $q$ in time interval $t$. 
Since it is plausible to assume that initially popular
topics will stay popular later on in time as well, we can calculate the ratios
$C_q(t_i, t_j) = N_q(t_i) / N_q(t_j)$ for topic $q$ for time frames $t_i$ and
$t_j$. Figure~\ref{fig:count_densities}(a) shows the distribution of $C_q(t_i,
t_j)$'s over all topics for four arbitrarily chosen pairs of time frames
(nevertheless such that $t_i > t_j$, and $t_i$ is relatively large, and $t_j$ is
small).

These figures immediately suggest that the ratios $C_q(t_i, t_j)$ are
distributed according to log-normal distributions, since the horizontal axes are
logarithmically rescaled, and the histograms appear to be Gaussian functions. To
check if this assumption holds, consider Fig.~\ref{fig:count_densities}(b),
where we show the Q-Q plots of the distributions of
Fig.~\ref{fig:count_densities}(a) in comparison to normal distributions. We can
observe that the (logarithmically rescaled) empirical distributions exhibit
normality to a high degree for later time frames, with the exception of the high
end of the distributions. These 10-15 outliers occur more frequently than could
be expected for a normal distribution.

Log-normals  arise as a result of  multiplicative growth
processes with noise~\cite{Mitzenmacher2004Brief}. In our
case, if $N_q(t)$ is the number of tweets for a given topic $q$ at time $t$,
then the dynamics that leads to a log-normally distributed $N_q(t)$ over $q$ can
be written as:
\begin{equation}
N_q(t) = \left[ 1 + \gamma(t) \xi(t) \right] N_q(t - 1),
\label{eq:multiplicative_growth}
\end{equation}
where the random variables $\xi(t)$ are positive, independent and identically
distributed as a function of $t$ with mean $1$ and variance $\sigma^{2}$. Note
that time here is measured in discrete steps ($t - 1$ expresses the previous
time step with respect to $t$), in accordance with our measurement setup.
$\gamma(t)$ is introduced to account for the novelty decay~\cite{Wu07}. We would expect topics to 
initially increase in popularity but to slow down their activity as they become obsolete or 
known to most users. Since $\gamma(t)$ is made up
of decreasing positive numbers, the growth of $N_{t}$ slows with time.

To see that Eq.~(\ref{eq:multiplicative_growth}) leads to a log-normal
distribution of $N_q(t)$, we first expand the recursion relation:
\begin{equation}
N_q(t) = \prod_{s = 1} ^ {t} \left[ 1 + \gamma(s) \xi(s) \right] N_q(0).
\label{eq:Nq_expressed}
\end{equation}
Here $N_q(0)$ is the initial number of tweets in the earliest time step.
Taking the logarithm of both sides of Eq.~(\ref{eq:Nq_expressed}),
\begin{equation}
\ln N_q(t) - \ln N_q(0) = \sum_{s = 1} ^ {t} \ln \left[ 1 + \gamma(s) \xi(s)
\right]
\label{eq:summed_noise_terms}
\end{equation}
The RHS of Eq.~(\ref{eq:summed_noise_terms}) is the sum of a large number of
random variables. The central limit theorem states thus that if the random
variables are independent and identically distributed, then the sum asymptotically
approximates a normal distribution. The i.i.d condition would hold exactly for the
$\xi(s)$ term, and it can be shown that in the presence of the discounting
factors (if the rate of decline is not too fast), the resulting distribution is
still normal~\cite{Wu07}.

In other words, we expect from this model that $\ln \left[ N_q(t) / N_q(0)
\right]$ will be distributed normally over $q$ when fixing $t$. These quantities
were shown in Fig.~\ref{fig:count_densities} above. Essentially, if the
difference between the two times where we take the ratio is big enough, the
log-normal property is observed.

The intuitive explanation for the multiplicative model of
Eq.~(\ref{eq:multiplicative_growth}) is that at each time step the number of
\emph{new} tweets on a topic is a multiple of the tweets that we already have.
The number of past tweets, in turn, is a proxy for the number of users that are
aware of the topic up to that point. These  users  discuss the topic on
different forums, including Twitter, essentially creating an effective network
through which the topic spreads. As more users talk about a particular topic, many others are likely to
learn about it, thus giving the multiplicative nature of the spreading. The
noise term is necessary to account for the stochasticity of this
process. On the other hand, the monotically decreasing $\gamma(t)$ characterizes
the decay in timeliness and novelty of the topic as it slowly becomes obsolete
and known to most users, and guarantees that $N_q(t)$ does not grow
unbounded~\cite{Wu07}.

\begin{figure}
\includegraphics[width=\linewidth]{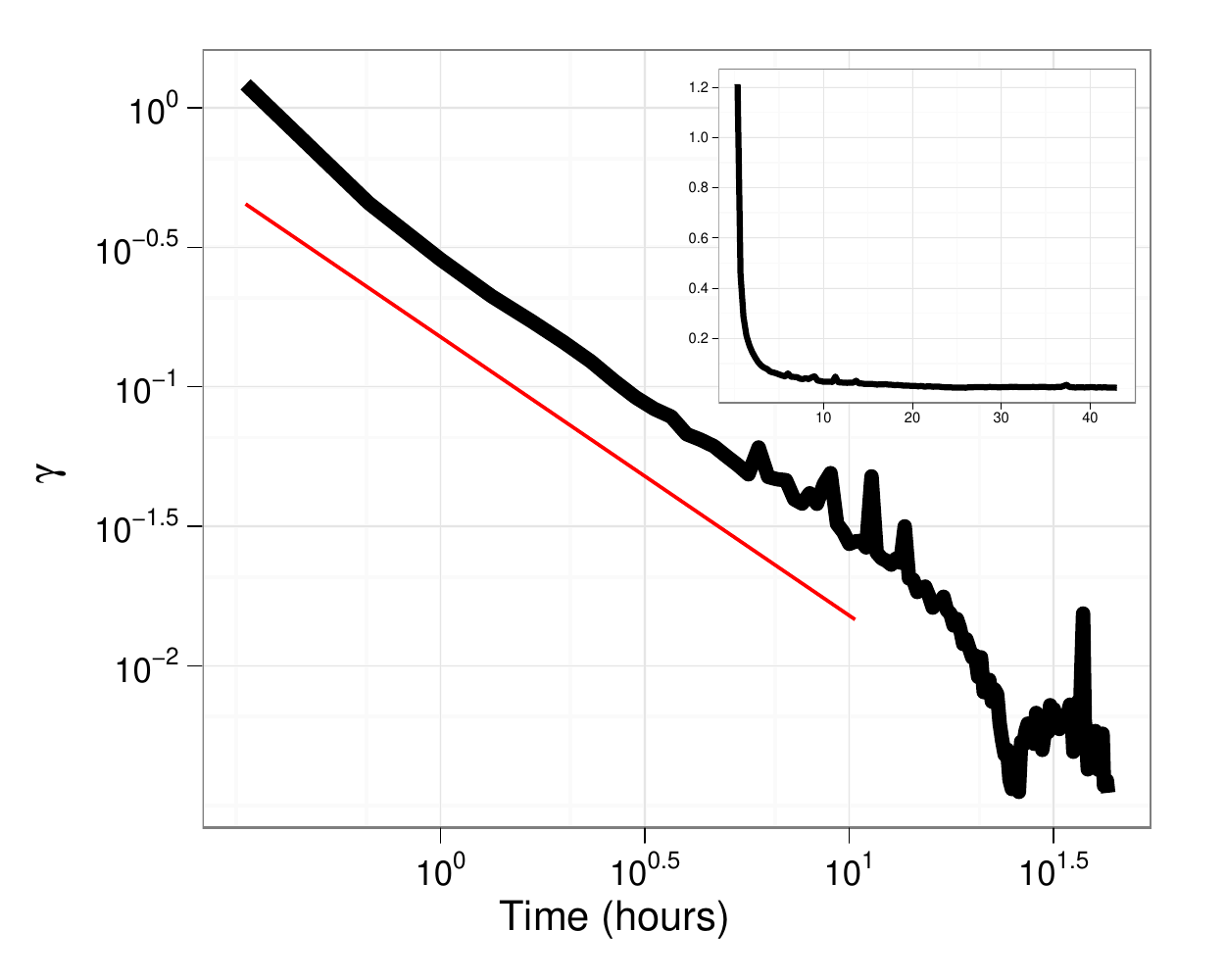}
\vspace{-0.1in}
\caption{{\bf The decay factor $\gamma(t)$ in time as measured using
Eq.~(\ref{eq:gamma_measurement}). The log-log plot exhibits that it decreases in
a power-law fashion, with an exponent that is measured to be exactly -1 (the
linear regression on the logarithmically transformed data fits with $R^2 =
0.98$). The fit to determine the exponent was performed in the range of the
solid line next to the function, which also shows the result of the fit while
being shifted lower for easy comparison. The inset displays the same $\gamma(t)$
function on standard linear scales.}}
\label{fig:gamma_in_time}
\end{figure}

To measure the functional form of $\gamma(t)$, we observe that the expected
value of the noise term $\xi(t)$ in Eq.~(\ref{eq:multiplicative_growth}) is
$1$. Thus averaging over the fractions between consecutive tweet counts yields
$\gamma(t)$:
\begin{equation}
\gamma(t) = \left \langle \frac{N_q(t)}{N_q(t - 1)} \right \rangle_q - 1.
\label{eq:gamma_measurement}
\end{equation}
The experimental values of $\gamma(t)$ in time are shown in
Fig.~\ref{fig:gamma_in_time}. It is interesting to notice that $\gamma(t)$
follows a power-law decay very precisely with an exponent of $-1$, which means
that $\gamma(t) \sim 1 / t$.

\section{The growth of tweets over time}

The interesting fact about the decay function $\gamma(t) = 1 / t$ is that it
results in a \emph{linear increase} in the total number of tweets for a topic
over time. To see this, we can again consider Eq.~(\ref{eq:summed_noise_terms}),
and approximate the discrete sum of random variables with an integral of the
operand of the sum, and substitute the noise term with its expectation value,
$\langle \xi(t) \rangle = 1$ as defined earlier (this is valid if $\gamma(t)$ is
changing slowly). These approximations yield the following:
\begin{equation}
\ln \frac{N_q(t)}{N_q(0)} \approx \int_{\tau = 0} ^ {t} \ln \left[ 1 +
\gamma(\tau) \right] d\tau \approx \int_{\tau = 0} ^ {t} \frac{1}{\tau} d\tau =
\ln t.
\end{equation}
In simplifying the logarithm above, we used the Taylor expansion of $\ln (1 + x)
\approx x$, for small $x$, and also used the fact that $\gamma(\tau) =
1 / \tau$ as we found experimentally earlier.

It can be immediately seen then that $N_q(t) \approx N_q(0) \, t$ for the range
of $t$ where $\gamma(t)$ is inversely proportional to $t$. In fact, it can be
easily proven that no functional form for $\gamma(t)$ would yield a linear
increase in $N_q(t)$ other than $\gamma(t) \sim 1 / t$ (assuming that the above
approximations are valid for the stochastic discrete case). This suggests that
the trending topics featured on Twitter increase their tweet counts linearly in
time, and their dynamics is captured by the multiplicative noise model we discussed above.

\begin{figure}
\includegraphics[width=\linewidth]{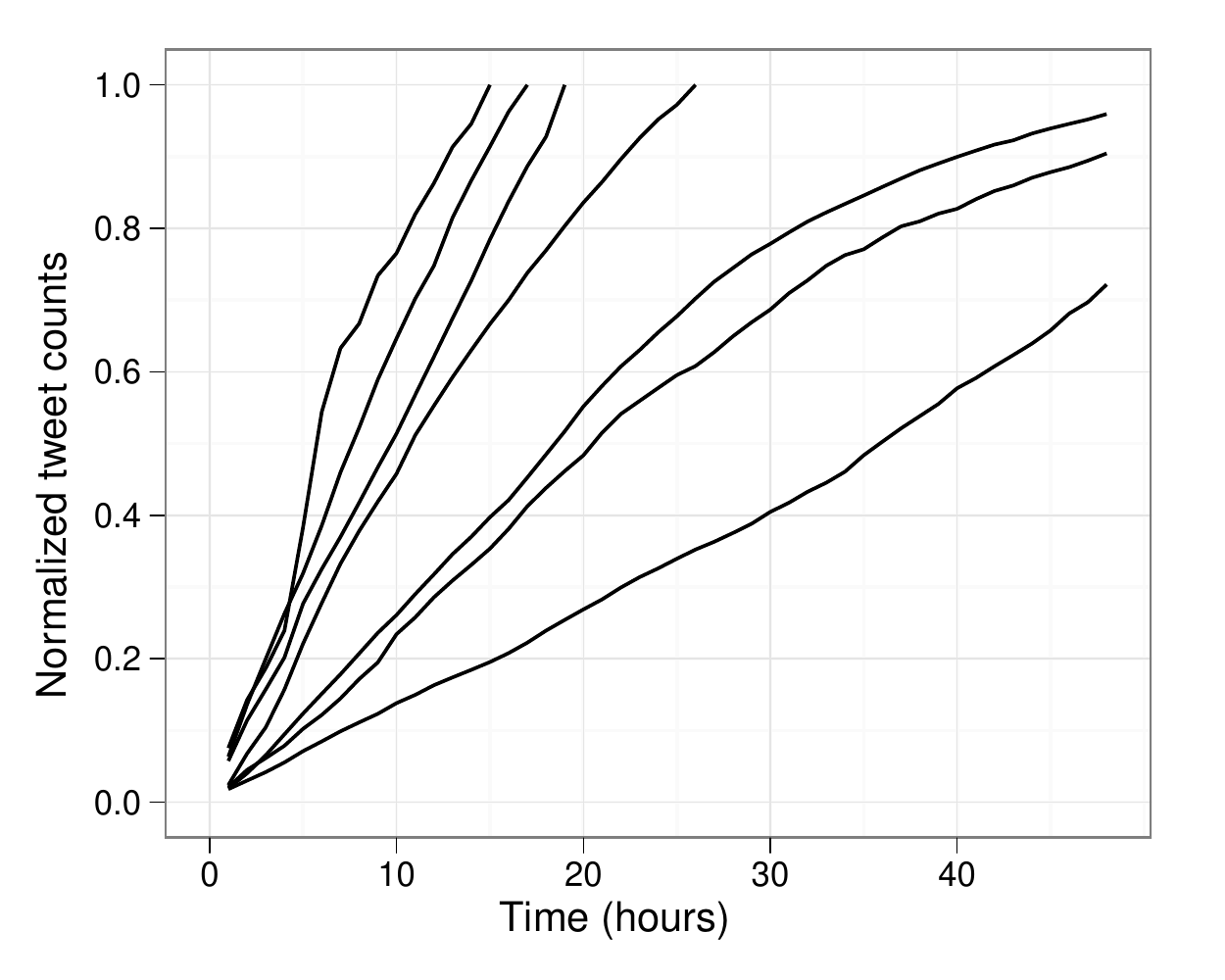}
\vspace{-0.24in}
\caption{{\bf The number of total tweets on topics in the first 48 hours,
normalized to $1$ so that they can be shown on the same plot. The randomly selected topics were (from
left to right): ``Earnings'', ``\#pulpopaul'', ``Sheen'', ``Deuces Remix'',
``Isaacs'', ``\#gmp24'', and ``Mac App''.}}
\label{fig:growth_examples}
\end{figure}

\begin{figure}
\includegraphics[width=\linewidth]{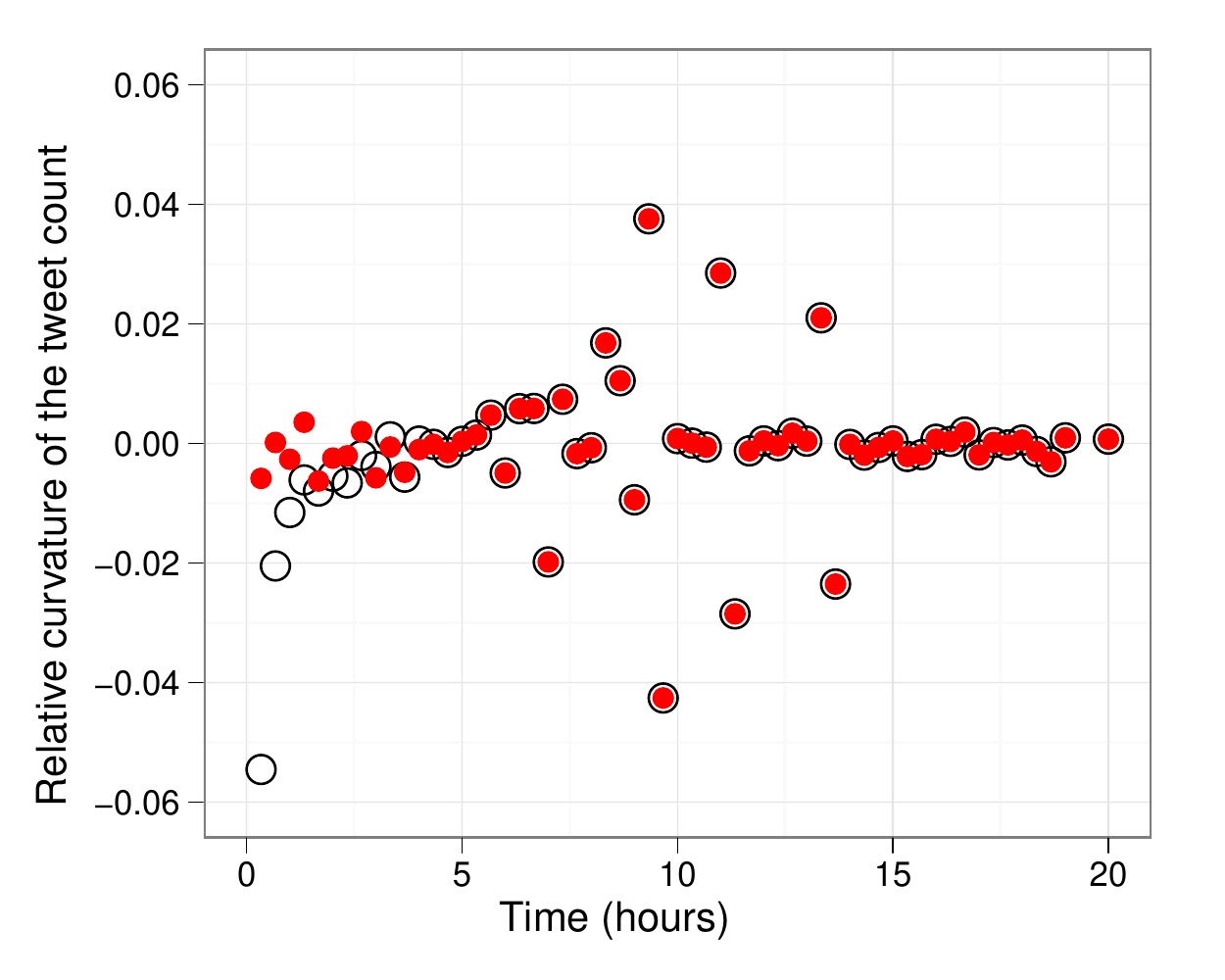}
\vspace{-0.24in}
\caption{{\bf The average of the second derivative of the total number of tweets
over all topics. For one topic, we first divided the derivative values by the
mean of the tweet counts so as to minimize the differences between the wide
range of topic popularities. The open circles show the derivatives obtained
with this procedure for all topics, while the smaller red dots represent only
topics that trended for longer than 4 hours.}}
\label{fig:diff_curvature}
\end{figure}

To check this, we first plotted a few representative examples of the
cumulative number of tweets for a few topics in Fig.~\ref{fig:growth_examples}.
It is apparent that all the topics ( selected randomly) show
an approximate initial linear growth in the number of tweets.We also
checked if this is true in general. Figure~\ref{fig:diff_curvature} shows the
second discrete derivative of the total number of tweets, which we expect to be
$0$ if the trend lines are linear on average. A positive second derivative
would mean that the growth is superlinear, while a negative one suggests that it
is sublinear. We point out that before taking the average of all second derivatives over the
different topics in time, we divided the derivatives by the average of the
total number of tweets of the given topics. We did this so as to account for the large
difference between the ranges of the number of tweets across topics, since a
simple averaging without prior normalization would likely bias the results
towards topics with large tweet counts and their fluctuations. The averages
are shown in Fig.~\ref{fig:diff_curvature}.

We  observe from the figure that when we consider all topics there is a
very slight sublinear growth regime right after the topic starts trending, which then
becomes mostly linear, as the derivatives data is distributed around $0$.
If we consider only very popular topics (that were on the trends site
for more than 4 hours),  we observe an even better linear trend. One reason for
this may be that topics that trend only for short periods exhibit a concave curvature, since
they lose popularity quickly, and are removed from among the Twitter trends by
the system early on.

These results suggest that once a topic is highlighted as a trend on a very visible
website, its growth becomes linear in time. The reason for this may be that as
more and more visitors come to the site and see the trending topics there is a
constant probability that they will also talk and tweet about it. This is in
contrast to scenarios where the primary channel of information flow is more
informal. In that case we expect that the growth will exhibit first a
phase with accelerated growth and then slow down to a point when no one
talks about the topic any more. Content that spreads through a social network
or without external ``driving'' will follow such a course, as has been showed
elsewhere~\cite{Szabo2010Predicting,Yang2011}.

\section{Persistence of Trends}

\begin{figure}[h]

 \includegraphics[height=6cm,width=7.0cm]{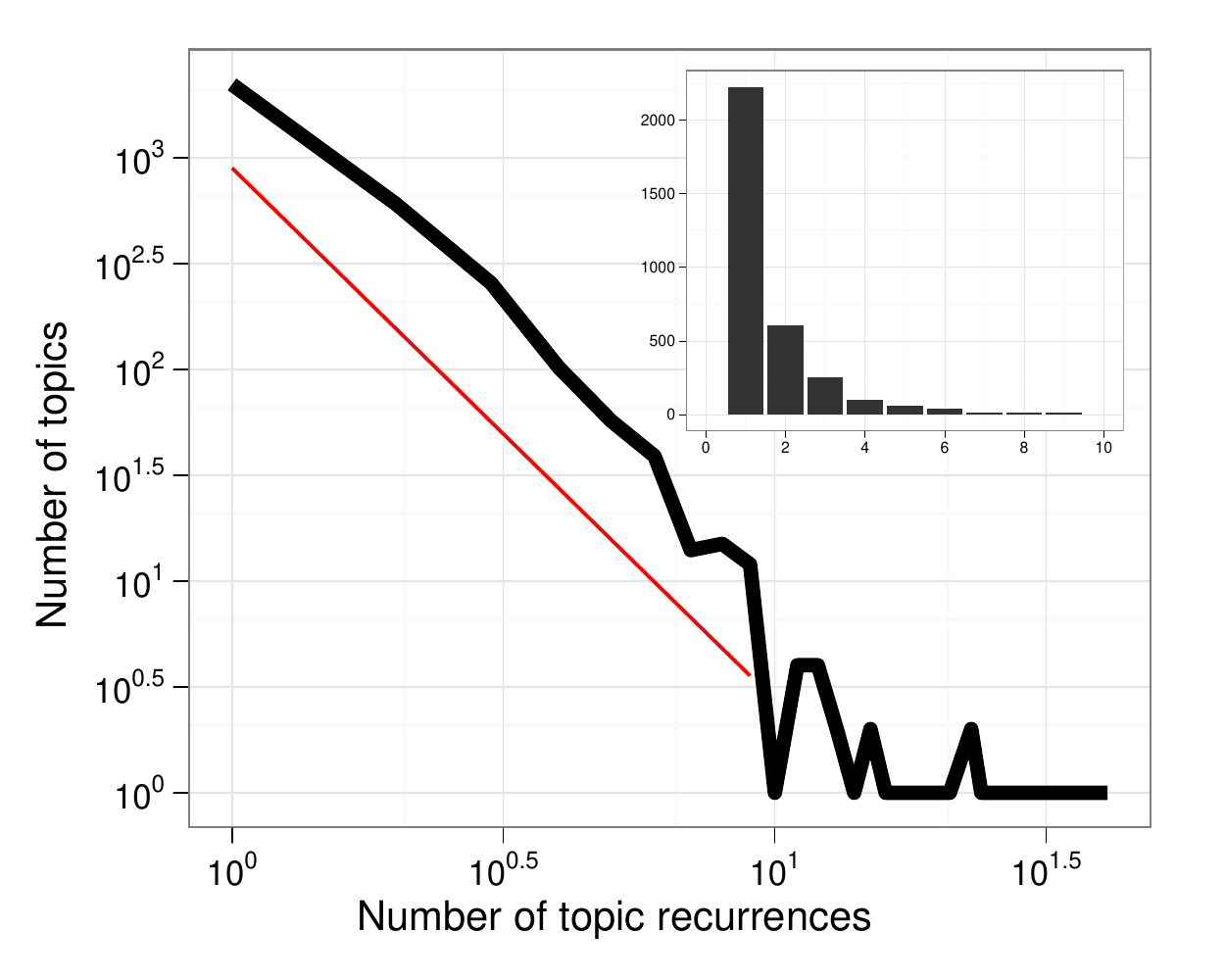} \\
 \includegraphics[height=6cm,width=7.0cm]{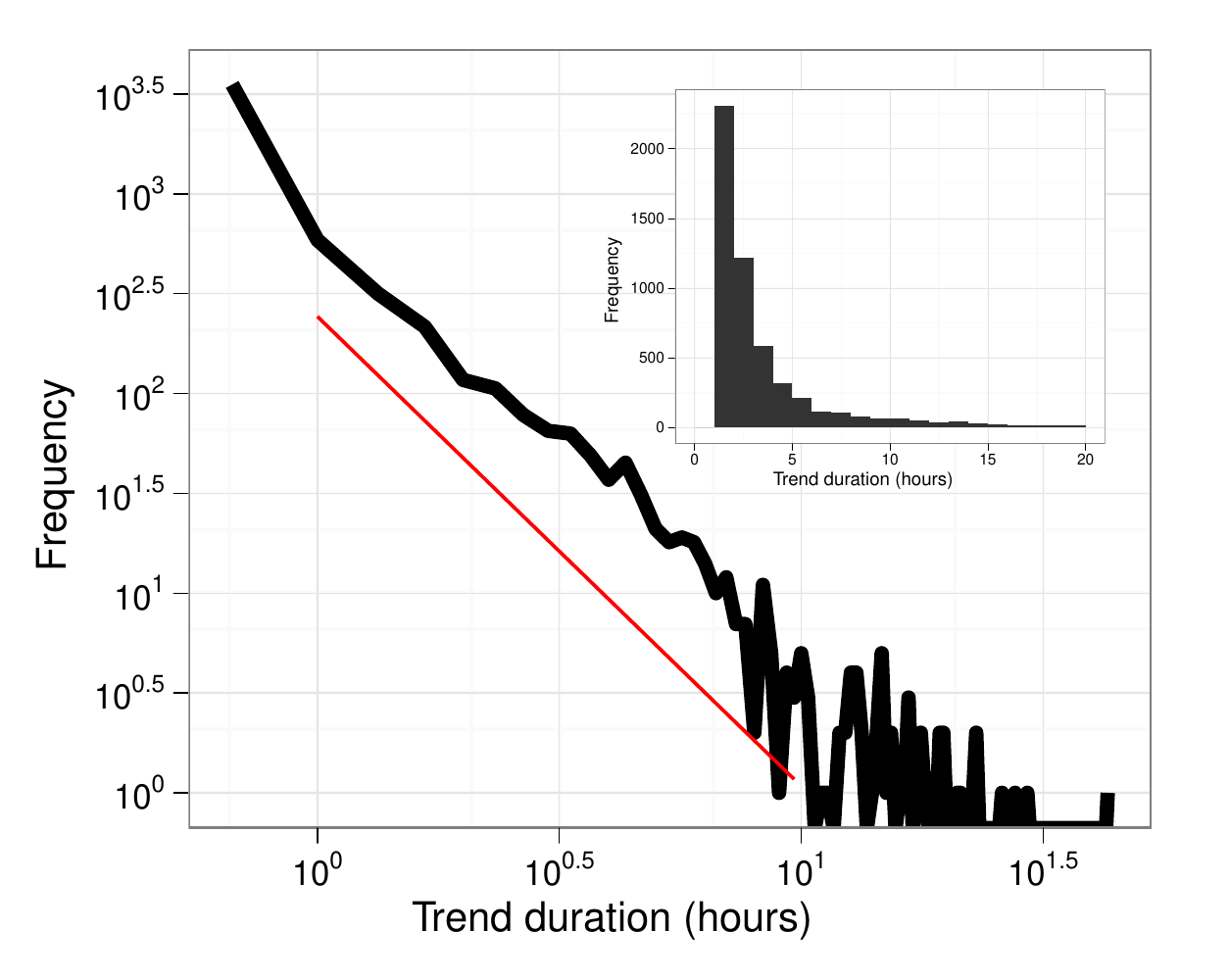}
\vspace{-0.1in}
\caption{{\bf \emph{(a)} The distribution of the number of sequences a trending topic comprises of \emph{(b)} The distribution of the lengths of each sequence. Both graphs are shown in the log-log scale with the inset giving the actual histograms in the linear scale.}}
\label{sequences}
\end{figure}

An important reason to study trending topics on Twitter is to
understand why some of them remain at the top while others dissipate quickly. To
see the general pattern of behavior on Twitter, we examined the lifetimes of the
topics that trended in our study. From Fig~\ref{sequences}(a) we can see that while
most topics occur continuously, around 34\% of topics appear in more than one
sequence. This means that they stop trending for a certain period of time before
beginning to trend again.

A reason for this behavior may be the time zones that are involved.
For instance, if a topic is a piece of news relevant to North American readers, a
trend may first appear in the Eastern time zone, and 3 hours later in the
Pacific time zone. Likewise, a trend may return the next morning if it was
trending the previous evening, when more users check their accounts again after
the night. 

Given that many topics do not occur continuously, we examined the distribution of the lengths sequences for all topics.
In Fig~\ref{sequences}(b) we show the length of the topic sequences. It can be observed 
that this is a power-law which means that most topic sequences are short and a few topics last for a very long time.
This could be due to the fact that  there are many topics competing for
attention. Thus, the topics that make it to the top (the trend list) last for a short time.
However, in many cases, the topics return to trend for more time, which is captured by
the number of sequences shown in Fig~\ref{sequences}(a), as mentioned.

\subsection{Relation to authors and activity}

We first examine the authors who tweet about given trending topics to see if the
authors change over time or if it is the same people who keep tweeting to cause trends. When we computed the correlation in the number of unique authors for a topic with the duration (number of timestamps) that the topic trends we noticed that correlation is very strong (0.80). This indicates that as the number of authors increases so does the lifetime, suggesting that the propagation through the network causes the topic to trend.
\begin{figure}[h]
 \centering
 \includegraphics[height=6cm,width=7.0cm]{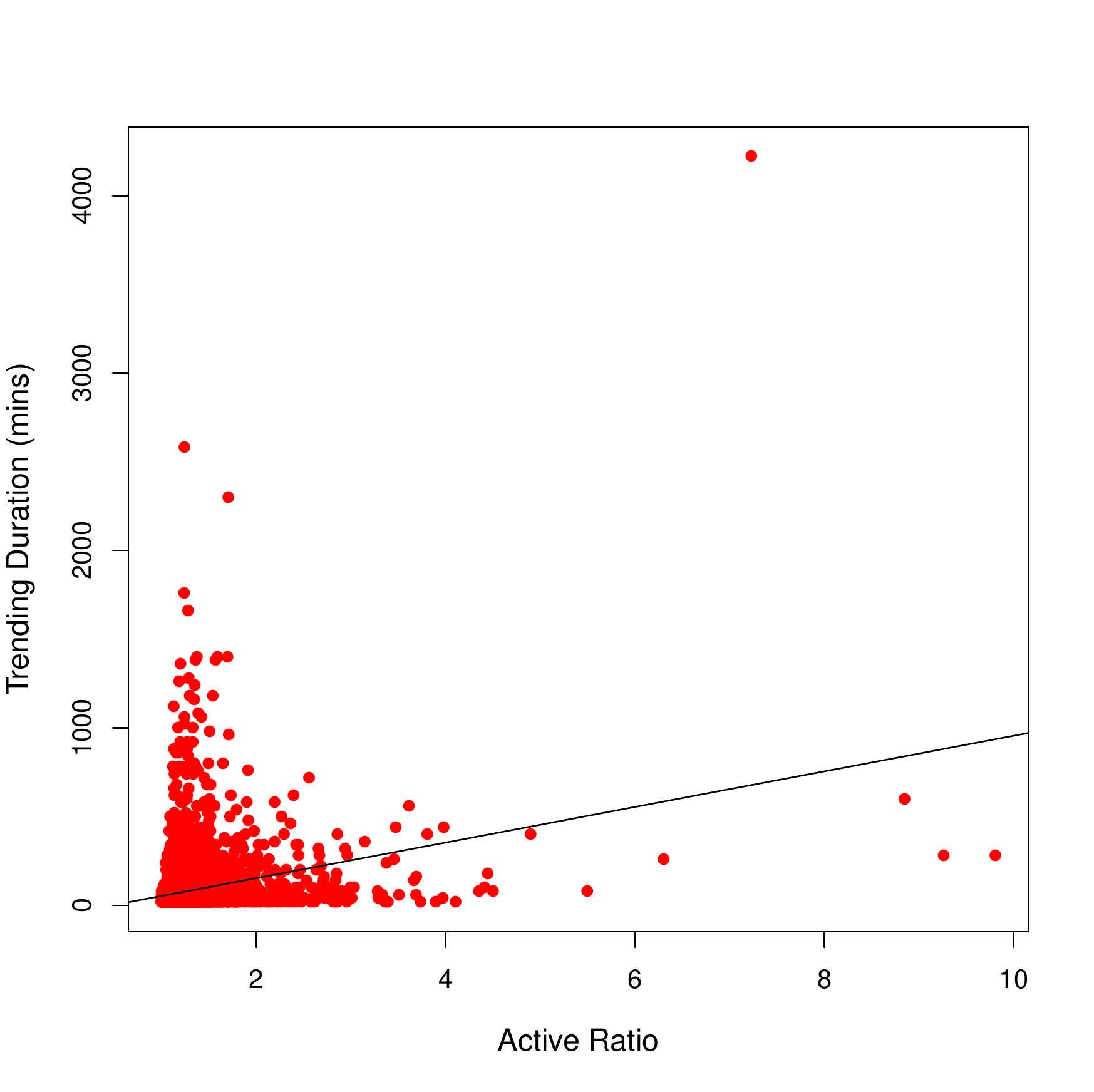} 
\vspace{-0.1in} 
\caption{{\bf Relation between the active-ratio and the length of the trend across all topics, showing that the active-ratio does not vary significantly with time.}}
\label{active}
\end{figure}

To measure the impact of authors we compute for each
topic the active-ratio $a_q$ as:
\begin{equation}
a_q=\frac{Number\ of\ Tweets}{Number\ of\ Unique\ Authors}
\end{equation}
The correlation of active-ratio with trending duration is as shown in
Fig~\ref{active}.
We observe that the
active-ratio quickly saturates and varies little with time for
any given topic. Since the authors change over time with the topic propagation,
the correlation between number of tweets and authors is high (0.83).	
\subsection{Persistence of long trending topics}

\begin{figure}[tbp]
 \centering
 \includegraphics[height=6cm,width=7.0cm]{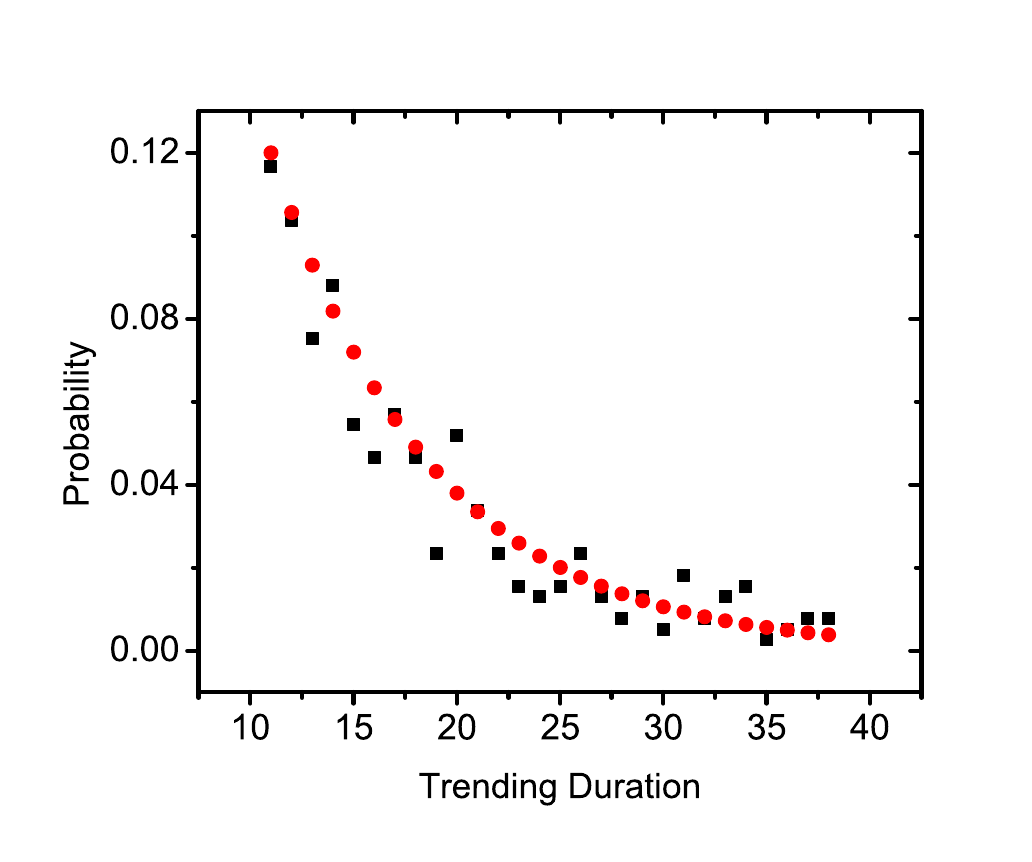}
\vspace{-0.1in} 
\caption{{\bf Distribution of trending times. The black dots represents actual
 trending data pulled from Twitter, and the red dots are the predictions from a geometric distribution
 with p=0.12.}}
\label{geom1}
\end{figure}
\begin{figure}[h]
 \centering
 \includegraphics[height=6cm,width=7.0cm]{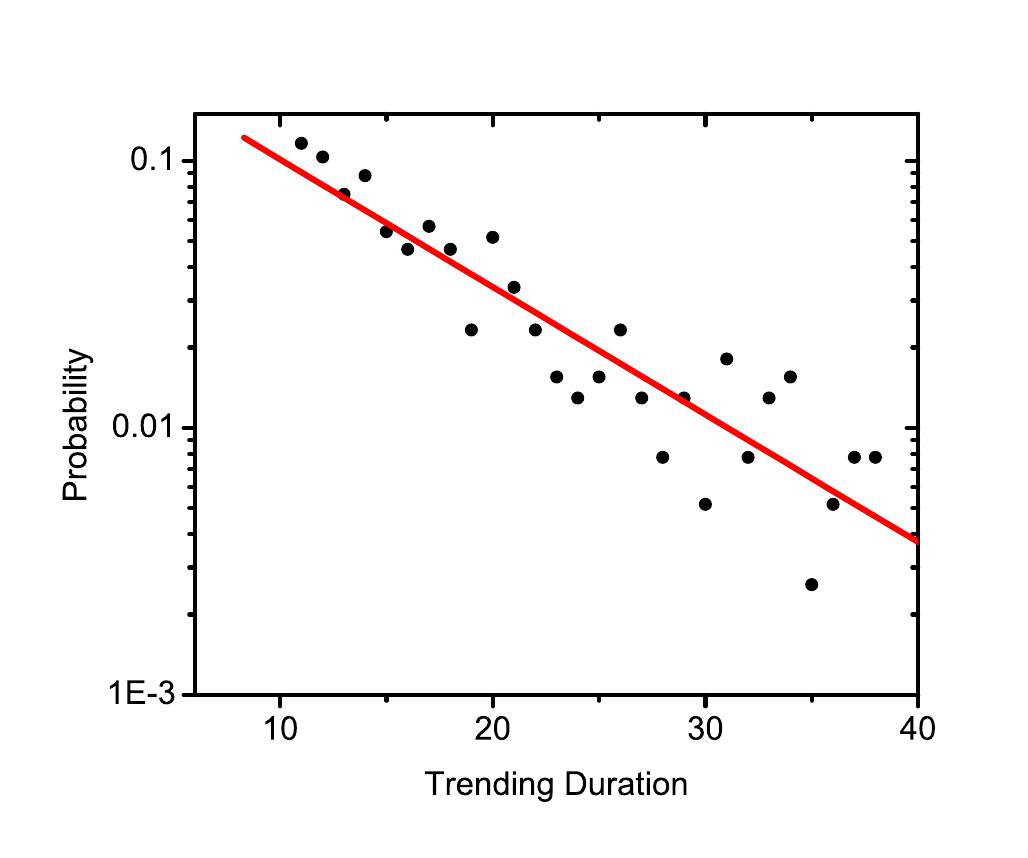}
\vspace{-0.1in} 
\caption{{\bf Fit of trending duration to density in log scale. The straight line suggests
 an exponential family of the trending time distribution. The red line gives a fit with an $R^{2}$ of 0.9112.}}
\label{geom2}
\vspace{-0.14in}
\end{figure} 
On Twitter  each topic competes with the others to 
survive on the trending page. As we now show, for the long trending ones we can derive
an expression for the distribution of their average length.  

We assume that, if the relative growth rate of tweets, denoted by $\phi_t = \frac{N_t}{N_{t-1}}$, 
falls below a certain threshold $\theta$, the topic would stop trending. 
When we consider long-trending topics, as they grow in time, they overcome the initial novelty decay, and the $\gamma$ term in equation (3) becomes fairly constant.
So we can measure the change over time using only the random variable $\xi$ as :

\begin{equation}
\log\phi_t = \log \frac{N_t}{N_{t-1}} =\log \frac{N_t}{N_{0}}-\log
\frac{N_{t-1}}{N_0} \simeq \xi_t
\end{equation}

Since the $\xi_s$ are independent and identical distributed
random variables, $\phi_1,\phi_2,\cdot\cdot\cdot\phi_t$ would be
independent with each other. Thus the probability that a topic stops
trending in a time interval $s$, where $s$ is large, is equal to the probability that
$\phi_s$ is lower than the threshold $\theta$, which can be written
as:
\begin{equation}
\begin{aligned}
p = \Pr(\phi_s<\theta)=\Pr(\log\phi_s<\log(\theta))\\
= \Pr(\xi_s<\log(\theta))= F(\log\theta)
\end{aligned}
\end{equation}

$F(x)$ is the cumulative distribution function of the random variable
$\chi$. Given that distribution we
can actually determine the threshold for survival as:
\begin{equation}
\theta  = e^{F ^{ - 1} (p)}
\end{equation}
From the independence property of the $\phi$, the duration or life time
of a trending topic, denoted by $L$,  follows a geometric
distribution, which in the continuum case becomes the exponential distribution.
Thus, the probability that a topic survives in the first $k$ time intervals and fails in the
$k+1$ time interval, given that $k$ is large, can be written as:

\begin{equation}
\Pr(L=k)=(1-p)^{k}p
\end{equation}

The expected length of trending duration $L$ would thus be:
\begin{equation}
<L>  = \sum\limits_0^\infty {(1 - p)^k p \cdot k} = \frac{1}{p} -
1=\frac{1}{F(\log\theta)}-1
\end{equation}

We considered trending durations for topics that trended for more than 10 timestamps on Twitter.
The comparison between the geometric distribution and the trending duration is
shown in Fig~\ref{geom1}. In Fig~\ref{geom2} the fit of the trending duration
to density in a logarithmic scale suggests an exponential function for the trending time. The R-square of the fitting
 is 0.9112. 

\section{Trend-setters}

We consider two types of people who contribute to trending topics - the sources who begin trends, and the propagators 
who are responsible for those trends propagating through the network due to the nature of the content they share.
\begin{figure}[h]
\vspace{-0.1in}
\includegraphics[height=6cm,width=7.0cm]{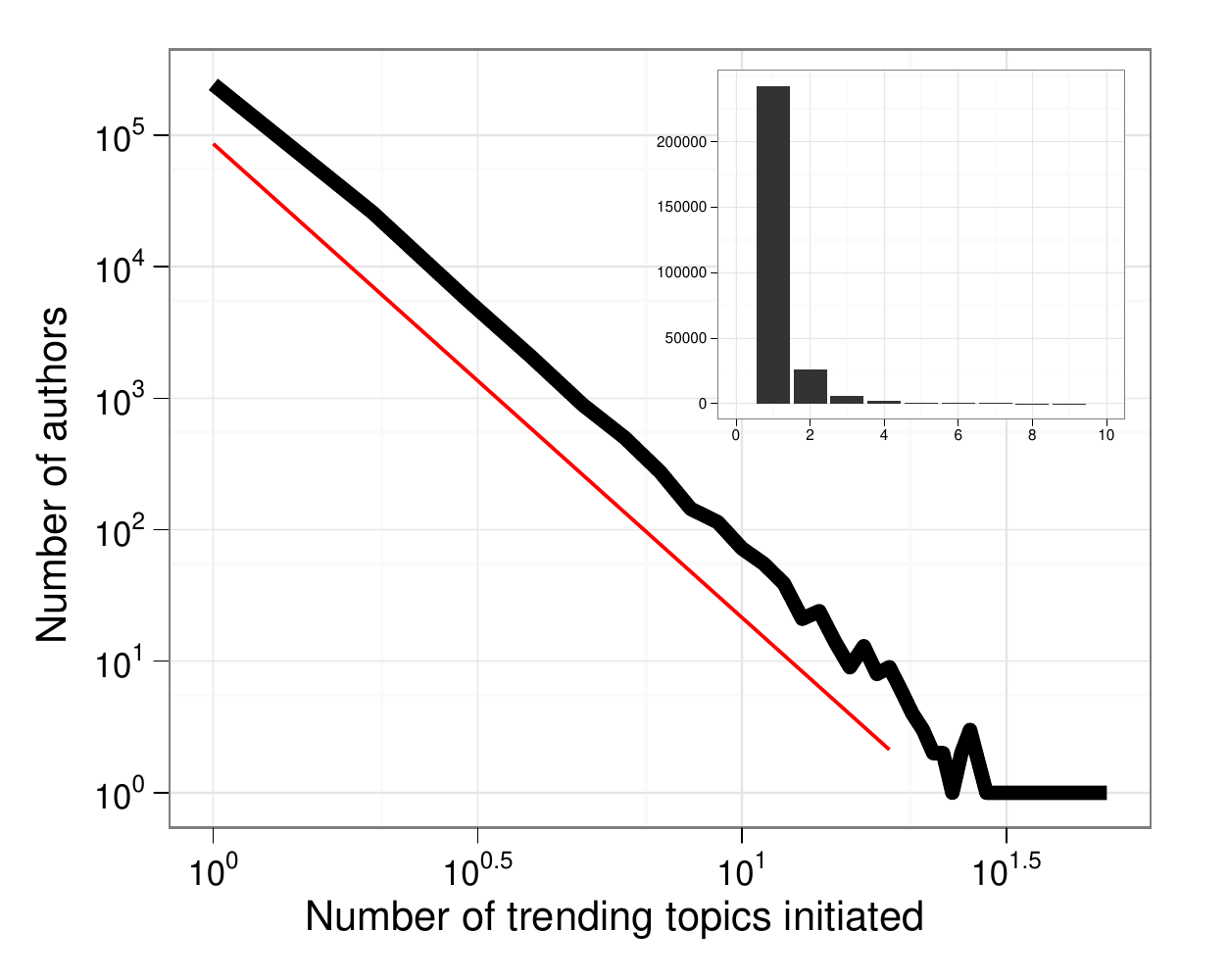}
\vspace{-0.14in}
\caption{{\bf Distribution of the first 100 authors for each trending topic. The log-log plot shows a power-law distribution. The inset graph gives the actual histogram in the linear scale.}}
\label{auth_top}
\vspace{-0.1in}
\end{figure}

\subsection{Sources}
We examined the users who initiate the most trending topics. 
First, for each topic we extracted the first 100 users who tweeted about it prior to its trending.
The distribution of these authors and the topics is a power-law, as shown in
Fig~\ref{auth_top}. This shows that there are few authors who contribute to
the creation of many different topics. To focus on these multi-tasking users, we considered only the authors who contributed to at least five trending topics.

When we consider people who are influential in starting trends on Twitter, we
can hypothesize two attributes - a high frequency of activity for these users, as well as a large follower
network. To evaluate these hypotheses we measured these two attributes for these
authors over these months. 

{\bf Frequency:} The tweet-rate can effectively measure the frequency of participation of a Twitter user.
The mean tweet-rate for these users was $26.38$ tweets per day,
indicating that these authors tweeted fairly regularly. However, when
we computed the correlation of the tweet-rate with the number of trending topics that
they contributed to, the result was a weak positive correlation of 0.22. This
indicates that although people who tweet a lot do tend to contribute to the
trending topics, the rate by itself does not strongly determine the popularity
of the topic. In fact, they happen to tweet on a variety of topics, many of which do not 
become trends. We found that a large number of them tended to tweet frequently about sporting events and players and teams involved. When some sports-related topics begin to trend, these users are among the early initiators of them, by virtue of their high tweet-rate. This suggests that the nature of the content plays a strong role in determining if a topic trends, rather than the users who initate it.

{\bf Audience:} When we looked at the number of followers for these authors, we were
surprised to find that they were almost completely uncorrelated (correlation of 0.01) with the number of trending topics, 
although the mean is fairly high (2481)~\footnote{This is due to the
fact that one of these authors has more than a million followers}. The
absence of correlation indicates that the number of followers is not an
indication of influence, similar to observations in earlier work~\cite{Romero2011}.

\subsection{Propagators}

We have observed previously that topics trend on Twitter mainly due to the propagation 
through the network. The main way to propagate information on Twitter is by retweeting. 
31\% of the tweets of trending topics are retweets. This reflects a high volume of propagation 
that garner popularity for these topics. Further, the number of retweets for a topic correlates very 
strongly (0.96) with the trend duration, indicating that a topic is of interest as long as there are people 
retweeting it. 

Each retweet credits the original poster of the tweet. Hence, to identify the authors who are retweeted the most in the trending topics, we counted the 
number of retweets for each author on each topic.
\begin{table}
\centering
\begin{tabular}{c|c|c|c}
\hline
Author & Retweets & Topics & Retweet-Ratio\\
\hline
vovo\_panico & 11688 & 65 & 179.81 \\
cnnbrk & 8444 & 84 & 100.52 \\
keshasuja & 5110 & 51 & 100.19\\
LadyGonga & 4580 & 54 & 84.81\\
BreakingNews & 8406 & 100 & 84.06\\
MLB & 3866 & 62 & 62.35\\
nytimes & 2960 & 59 & 50.17\\
HerbertFromFG & 2693 & 58 & 46.43\\
espn & 2371 & 66 & 35.92\\
globovision & 2668 & 75 & 35.57\\
huffingtonpost & 2135 & 63 & 33.88\\
skynewsbreak & 1664 & 52 & 32\\
el\_pais & 1623 & 52 & 31.21\\
stcom & 1255 & 51 & 24.60\\
la\_patilla & 1273 & 65 & 19.58\\
reuters & 957 & 57 & 16.78\\
WashingtonPost & 929 & 60 & 15.48\\
bbcworld & 832 & 59 & 14.10\\
CBSnews & 547 & 56 & 9.76\\
TelegraphNews & 464 & 79 & 5.87\\
tweetmeme & 342 & 97 & 3.52\\
nydailynews & 173 & 51 & 3.39\\
\hline
\end{tabular}
\vspace{-0.1in}
\caption{{\bf Top 22 Retweeted Users in at least 50 trending topics each}}
\vspace{-0.2in}
\label{infs}
\end{table}

{\bf Domination:} We found that in some cases, almost all the retweets for a topic are credited to 
one single user. These are topics that are entirely based on the comments by that user. They can thus be said 
to be dominating the topic. The \emph{domination-ratio} for a topic can be defined as the fraction of the retweets of that topic that can be attributed to the largest contributing user for that topic. However, we observed a negative correlation of $-0.19$ between the domination-ratio of a topic to its trending duration. This means that topics revolving around a 
particular author's tweets do not typically last long. This is consistent with the earlier observed strong correlation between number of authors and the trend duration. Hence, for a topic to trend for a long time, it requires many people to contribute actively to it.

{\bf Influence:} On the other hand, we observed that there were authors who contributed actively to 
many topics and were retweeted significantly in many of them. For each author, we computed the ratio of 
retweets to topics which we call the \emph{retweet-ratio}. The list of influential authors who are retweeted in at least 50 
trending topics is shown in Table~\ref{infs}. We find that a large portion of these authors are popular news sources such 
as CNN, the New York Times and ESPN. This illustrates that social media, far from being an alternate source of news, functions more as a filter and an amplifier for interesting news from traditional media.

\section{Conclusions}
To study the dynamics of trends in social media, we have conducted a comprehensive study on trending topics on Twitter. We first
derived a stochastic model to explain the growth of trending topics and showed that it leads to a lognormal distribution, which is validated by our empirical results. We also have found that most topics do not trend for long, and for those that are long-trending, their persistence obeys a geometric distribution.

When we considered the impact of the users of the network, we discovered that the number of followers and tweet-rate of users are not the attributes that cause trends. What proves to be more important in determining trends is the retweets by other users, which is more related to the content that is being shared than the attributes of the users. Furthermore, we found that the content that trended was largely news from traditional media sources, which are then amplified by repeated retweets on Twitter to generate trends. 
\bibliographystyle{habbrv}

\end{document}